\documentclass[conference,a4paper]{IEEEtran}
\IEEEoverridecommandlockouts

\usepackage[utf8]{inputenc} 
\usepackage[T1]{fontenc}
\usepackage{url}
\usepackage{ifthen}
\usepackage{cite}
\usepackage[cmex10]{amsmath}

\usepackage{graphicx}
\usepackage{amsfonts}
\usepackage{amssymb}
\usepackage{mathrsfs}
\usepackage{enumitem}
\usepackage{subfigure}
\usepackage{mdwmath}
\usepackage{mdwtab}
\usepackage{dsfont}
\usepackage[ruled,vlined,linesnumbered]{algorithm2e} 

\newtheorem{remark}{Remark}

\def\C{{\boldsymbol{C}}}
\def\D{{\boldsymbol{D}}}

\def\I{{\boldsymbol{I}}}
\def\J{{\boldsymbol{J}}}

\def\e{{\boldsymbol{e}}}

\def\h{{\boldsymbol{h}}}

\def\m{{\boldsymbol{m}}}
\def\n{{\boldsymbol{n}}}

\def\s{{\boldsymbol{s}}}

\def\u{{\boldsymbol{u}}}
\def\v{{\boldsymbol{v}}}

\def\x{{\boldsymbol{x}}}
\def\y{{\boldsymbol{y}}}
\def\z{{\boldsymbol{z}}}
\newenvironment{iarray}{\begin{IEEEeqnarray}{rCl}}{\end{IEEEeqnarray}\ignorespacesafterend}
\newcommand{\dx}{{\sf d}}

\begin{document}
	
	\title{
		Inferring Remote Channel State Information: Cram\'er-Rao Lower Bound and Deep Learning Implementation
	}
	
	\author{\IEEEauthorblockN{Zhiyuan Jiang$^1$, Ziyan He$^1$, Sheng Chen$^1$, Andreas F. Molisch$^2$, \IEEEmembership{Fellow,~IEEE},\\ Sheng Zhou$^1$,  Zhisheng Niu$^1$, \IEEEmembership{Fellow,~IEEE}
		}
		\IEEEauthorblockA{$^1$\{zhiyuan@, hezy14@mails., chen-s16@mails., sheng.zhou@, niuzhs@\}tsinghua.edu.cn, Tsinghua University, Beijing, China\\
		   $^2$molisch@usc.edu, University of Southern California, Los Angeles, USA}
	}
	
	\maketitle
	
	\begin{abstract}
		Channel state information (CSI) is of vital importance in wireless communication systems. Existing CSI acquisition methods usually rely on pilot transmissions, and geographically separated base stations (BSs) with non-correlated CSI need to be assigned with orthogonal pilots which occupy excessive system resources. Our previous work adopts a data-driven deep learning based approach which leverages the CSI at a local BS to infer the CSI remotely, however the relevance of CSI between separated BSs is not specified explicitly. In this paper, we exploit a model-based methodology to derive the Cram\'er-Rao lower bound (CRLB) of remote CSI inference given the local CSI. Although the model is simplified, the derived CRLB explicitly illustrates the relationship between the inference performance and several key system parameters, e.g., terminal distance and antenna array size. In particular, it shows that by leveraging multiple local BSs, the inference error exhibits a larger power-law decay rate (w.r.t. number of antennas), compared with a single local BS; this explains and validates our findings in evaluating the deep-neural-network-based (DNN-based) CSI inference. We further improve on the DNN-based method by employing dropout and deeper networks, and show an inference performance of approximately $90\%$ accuracy in a realistic scenario with CSI generated by a ray-tracing simulator.
	\end{abstract}
	
	\section{Introduction}
	\label{sec_intro}
	Channel state information (CSI) plays a pivotal role in wireless communication systems, especially with the wide adoption of massive multiple-input multiple-output (MIMO) technology in the 5G and beyond systems. The knowledge of CSI can facilitate, and is necessary for, beamforming, spatial multiplexing, user scheduling and spatial diversity.
	
	The acquisition of CSI usually relies on pilot transmissions, i.e., known signals transmitted only for probing the propagation channel. Broadly speaking, without considering pilots for particular usage such as demodulation and phase offset correction, the pilot-based CSI acquisition involves either uplink pilots utilizing channel reciprocity \cite{Marzetta10} or downlink pilots and uplink CSI feedback \cite{Jiang14}, both entailing a severe signalling overhead in the system design. Traditional methods for overhead-reduced CSI acquisition usually involves leveraging the CSI \emph{linear correlations} in spatial, time and frequency domains. The majority of the related works focus on the spatial domain correlation. By transforming the CSI to the angular domain where most of the angular bins carry negligible energy, the dimensionality of CSI is significantly reduced \cite{Adhikary13, Jiang14, Rao14,brady13,lin17}. The time domain correlation is often modeled by the Gauss-Markov process, and work has been done considering the heterogeneity of channel coherence among users \cite{deng17}. There is also recent work \cite{had17} proposing a more efficient pilot packing which is enabled by orthogonal time frequency space (OTFS) modulation. In OFDM systems, the CSI correlation that can be exploited in the frequency domain stems from the sparse multi-path component (MPC) delay profile of the channel impulse response; such correlations can be exploited based on the compressive sensing framework \cite{gao16_cs}.
	
\begin{figure*}[!t]
    	\centering
    	\includegraphics[width=0.7\textwidth]{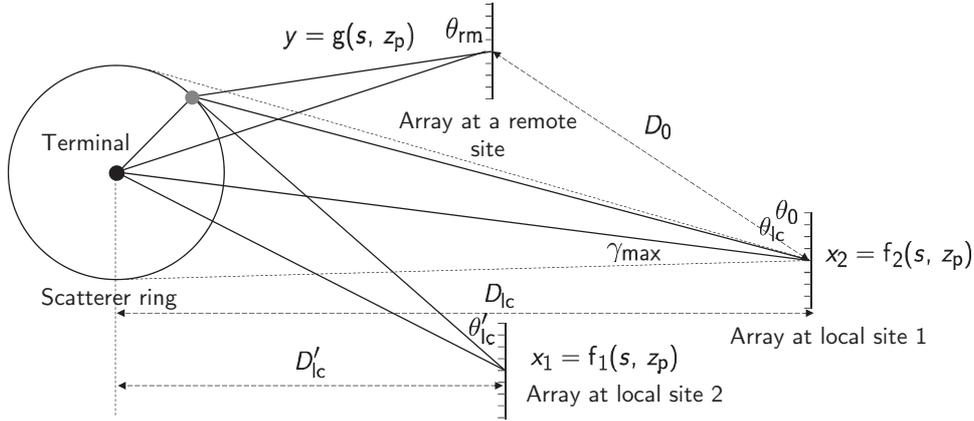}
    	\caption{A one-ring model based CSI inference scenario illustration. Two local sites with known CSI are described; the CSI at the remote site is unknown and to be inferred.}
    	\label{Fig_geo}
    \end{figure*}

	In the spatial domain, the linear correlation among co-located antennas proves helpful in reducing the CSI overhead; however, the distance between two remote base stations (BSs) renders this simple structure powerless since channels of remote sites (with distance much larger than wavelength, cf. Fig. \ref{Fig_geo}) are considered linearly independent. More sophisticated data structures in CSI, which are not obvious so far due to the complicated propagation environment, should be explored for CSI overhead reduction. Our previous work \cite{chen17,liu15} adopted a data-driven approach, particularly the deep neural network (DNN), to address this issue and showed promising results in inferring the CSI at a remote site based on observations at a local site. However, the work uses a completely model-free approach and hence no theoretical analysis was given.
    
	In this paper, a deeper understanding of the CSI structure is desired, towards which end we first describe the general remote CSI inference problem and accordingly present a model-based Cram\'er-Rao lower bound (CRLB) analysis. Although the employment of channel models, e.g., one-ring model \cite{lee73,Shiu00} and line-of-sight (LoS) model in this paper, or perhaps any other existing channel models in the literature, cannot fully describe the real-world propagation environment, they are useful in obtaining insightful and closed-form results. Afterwards, for realistic implementation we improve upon our deep learning based approach and achieve good and robust (w.r.t. propagation channels) performance. The main contributions include that we derive the closed-form CRLB of remote CSI inference when the channel is LoS; we find that the inference CRLB when considering a LoS and 2D scenario scales down as $1/M_\mathsf{lc}$ where $M_\mathsf{lc}$ denotes the number of local BS antennas when CSI of one local BS is used, and $1/M^3_\mathsf{lc}$ with CSI from more than one BSs; a novel DNN is proposed which exhibits improved performance compared with previous work, and more importantly, can be applied universally in real-world scenarios.
	
	\section{System Model and Problem Formulation}
	Consider a communication setup (see Fig. \ref{Fig_geo}) where the terminal (user equipment, UE) is surrounded by local scatterers. Let $\x$ be the CSI between the terminal and a number of BSs (henceforth called "local sites"). Let there be another BS, henceforth called ``remote site'', with CSI $\y$. The terminal has a single antenna element, while the BSs have antenna arrays of size $M_\mathsf{lc}$ for the local sites and $M_\mathsf{rm}$ for the remote site, respectively. The goal is to infer $\y$ from the knowledge of $\x$. Such an inference problem from $\x$ to $\y$ might be ill-posed, in the sense that any of the following conditions is not satisfied:
	\begin{itemize}
		\item 
		For any $\x$, there exists a unique $\y$ that corresponds to the channel realization.
		\item
		The mapping from $\x$ to $\y$ is steady, i.e., given a small error of $\x$, the corresponding error of $\y$ is also limited.
	\end{itemize}
	
	In practice, we can usually relax the first condition to that if the mapping is not unique, then the mapping error is, to some extent, acceptable. The inference problem is studied by considering the physical propagation environment. Define the propagation channel as a vector of parameters $\z_{\mathsf{p}}$, consisting of, e.g., scatterer locations, reflection attenuation factors and etc. Then both the CSIs at the local or remote site can be expressed as a function of $\z_{\mathsf{p}}$ and terminal location $\s$ based on the same methodology of ray-tracing models, i.e., 
	\begin{equation}
	\label{model}
	\x = \operatorname{f}(\s,\,\z_{\mathsf{p}})\textrm{ and }\y = \operatorname{g}(\s,\,\z_{\mathsf{p}}),
	\end{equation}
	where $\operatorname{f}(\cdot)$ and $\operatorname{g}(\cdot)$ denotes the function mappings from the physical environment and terminal location to the CSIs at the local sites and remote site, respectively. Based upon this, the inference problem is by and large solved if the physical environment and the terminal location are perfectly known. In particular, it has been shown \cite{molisch04_gscm,thomae03} that if a high-resolution estimation of the MPC parameters (angles, amplitudes) is available, then a prediction of the instantaneous channel impulse response can be made as long as the separation between local and remote sites is less than the stationarity region of the channel. However, in realistic situations we cannot invert the function mapping of $\x = \operatorname{f}(\s,\,\z_{\mathsf{p}})$ due to e.g., modelling errors, limited sampling points, insufficient observation accuracy and etc, rendering the whole inference problem ill-posed. 
	
	On the other hand, we argue that as $M_{\mathsf{lc}}$ increases, the inverse problem of $\x = \operatorname{f}(\s,\,\z_{\mathsf{p}})$ becomes less ill-conditioned. Furthermore, in the following sections we will show that with growing $M_{\mathsf{lc}}$, the inference problem can be addressed with more and more accuracy. This phenomenon is referred to as \emph{CSI manifestation}, as the propagation channel manifests from the CSI as the CSI spatial dimensionality grows. One of the main objectives of the paper is to obtain the scaling result of, e.g., the inference error, with $M_{\mathsf{lc}}$.\footnote{In this paper, we assume the uniform linear antenna arrays at the BS sites and  the system topology is two-dimensional.}
	
	\section{CRLB Analysis Based on One-Ring Channel Model}
	To gain insights into the channel inference problem and performance, as well as the CSI manifestation phenomenon, a theoretical analysis based on widely-adopted channel models is conducted in this section. The general model described in \eqref{model} is implicit and therefore infeasible for theoretical analysis. Towards this end, we adopt a model-based approach which essentially transforms the CSI inference problem to a parameter extraction problem based on a well-defined channel model; specifically, the one-ring ray-tracing channel model is used where the scatterers are assumed to be placed within a ring of radius $r_{\mathsf{max}}$ (only single-scattering is considered). The terminal is at the center of the scatterer-ring. The received signal $\y = [y_1,...,y_M]^{\mathsf{T}}$ at a site can be written by (denote the channel vector by $\h = [h_1,...,h_M]^{\mathsf{T}}$)
	\begin{iarray}
		\label{array_res}
		y_i &=& \sqrt{P_{\textrm{tx}}} h_i + n_i, \nonumber\\
		h_i &=& \sum_{k=1}^K g_{ki} \exp{\left(-\frac{j2\pi}{\lambda} d_{ki} \right)},
	\end{iarray}
	where $k$ is the index of MPC going through the $k$-th scatterer, $g_{ki} \triangleq \frac{\lambda}{4\pi d_{ki}}$ denotes the channel gain due to pathloss based on Friis' law, the Gaussian additive noise is denoted by $n_i$ with variance of $\sigma^2$ (the sounding signal is omitted for simplification and assumed with effective transmit power of $P_{\textrm{tx}}$), and the path distance of the $k$-th MPC received at the $i$-th antenna is denoted by $d_{ki}$ which equals
	\begin{equation}
	d_{ki} = \xi_{\mathsf{t}k} + \xi_{ki}
	\end{equation}
	with $\xi_{\mathsf{t}k}$ denoting the distance from the terminal to the $k$-th scatterer and $\xi_{ki}$ the distance from the $k$-th scatterer to the $i$-th receive antenna. Assuming that the angle spread at the terminal is relatively small and the terminal is in the far-field, i.e., $D \gg r_{\mathsf{max}}$ and $D \gg \frac{2 M^2_\mathsf{lc} \delta^2}{\lambda}$ where $\delta$ is antenna spacing \cite{abdi02}, the channel gains of all MPCs are therefore approximately identical, i.e., $g_{ki}=g$, $\forall 1\le k \le K, 1 \le i \le M$, and 
	\begin{equation}
	d_{ki} \approx \xi_k + i \delta \cos{\gamma_k}.
	\end{equation}
	Denote by $\xi_k$ the distance from the terminal to the antenna array (a reference point such that the above equation is upheld) passing through the $k$-th scatterer, and $\gamma_k$ is the angle-of-arrival (AoA) of the $k$-th MPC. Based on this approximation, \eqref{array_res} can be re-written as
	\begin{equation}
	\label{array_res_app}
	y_i = \sum_{k=1}^K g \sqrt{P_{\textrm{tx}}} \exp{\left(-\frac{j2\pi \delta i}{\lambda}  \cos{\gamma_k} + j \phi_k \right)} + n_i,
	\end{equation}
	where $\phi_k = -\frac{j2\pi}{\lambda} \xi_k$, and $g=\frac{\lambda}{4 \pi D}$ represents the pathloss.
	
	The angular power spectrum (APS) seen at a site is defined as 
	\begin{equation}
	S(\gamma) = g \mu(\gamma) p(\gamma),
	\end{equation}
	where $\gamma$ is the AoA to the site and $S(\gamma)$ is normalized such that
	\begin{equation}
	\int_{-\gamma_{\mathsf{max}}}^{\gamma_{\mathsf{max}}}  S^2(\gamma) \dx \gamma = g^2.
	\end{equation}
	The scatterer angular distribution is characterized by $\mu(\gamma)$, and the probability that the MPC with AoA $\gamma$ is observable (not blocked) at the site is $p(\gamma)$. For example, if we assume that the scatterers are continuously placed on the ring with radius $r_{\mathsf{max}}$, then \cite{fulg02}
	\begin{equation}
	\mu_{\mathsf{ring}}(\gamma) = \frac{2}{\sqrt{\gamma_{\mathsf{max}}^2-(\gamma-\theta)^2}},
	\end{equation}
	where $\theta-\gamma_{\mathsf{max}} \le \gamma \le \theta+\gamma_{\mathsf{max}}$ and $\theta$ is the mean AoA of the terminal and $\gamma_{\mathsf{max}}$ denotes the maximum angular spread. A typical form of $p(\gamma)$ is e.g., uniform on the disk with a radius of $r_{\mathsf{max}}$. The channel array response is written as
	\begin{iarray}
	\label{int_res}
	y_i &=& \int_{-\gamma_{\mathsf{max}}}^{\gamma_{\mathsf{max}}} \sqrt{P_{\textrm{tx}}} S(\gamma) \nonumber\\
	&& \times \exp{\left(-\frac{j2\pi \delta i}{\lambda}  \cos{(\gamma+\theta)} + j \phi_{\gamma} \right)} \dx \gamma + n_i.
	\end{iarray}
	Combing with \eqref{model}, we denote 
	\begin{iarray}
		\label{relation}
		h_{\mathsf{lc},i} &=& \operatorname{f}\left(\theta_{\mathsf{lc}}, S_{\mathsf{lc}}(\gamma), \phi_{\mathsf{lc},\gamma}, r_{\mathsf{max},\mathsf{lc}}\right), \nonumber\\
		h_{\mathsf{rm},i} &=& \operatorname{g}\left(\theta_{\mathsf{rm}}, S_{\mathsf{rm}}(\gamma), \phi_{\mathsf{rm},\gamma}, r_{\mathsf{max},\mathsf{rm}}\right),
	\end{iarray}
	where the subscripts $(\cdot)_\mathsf{lc}$ and $(\cdot)_\mathsf{rm}$ denote the local site and the remote site, respectively. The function mapping of $\operatorname{f}(\cdot)$ and $\operatorname{g}(\cdot)$ are substantiated by \eqref{int_res}. Note that the model of \eqref{relation} allows, e.g., different angular spreads, different visibility of scatterers, different APSs at the local site and the remote site, by distinct $r_{\mathsf{max}}$, $p_{\gamma}$, $\mu(\gamma)$, respectively. It is thus significantly more general than a model valid within a stationarity region, which assumes that only the phases of MPCs vary (due to phase shifts related to the different run lengths).
	
	Based on the above model, the channel inference task can be stated concretely below:
	\begin{iarray}
		\textbf{P1:}\qquad && \textrm{Estimate } \h_{\mathsf{rm}}, \textrm{ given } \h_{\mathsf{lc}}, \textrm{ s.t., Eq. } \eqref{relation} \textrm{ is satisfied.}\nonumber\\
	\end{iarray}
	\subsection{Parameter Extraction and Inference: CRLB Analysis}
	It is observed that the implicit channel inference problem in \eqref{model} is transformed to a parameter extraction and estimation problem in \textbf{P1}, based on the adopted channel models; this allows us to derive the CSI inference CRLB which indicates the CSI inference accuracy. Equivalently, the inverse of the CRLB, representing the Fisher information, indicates the amount of information that the local CSI carries about the remote CSI.
	
	To further simplify the model and focus on the main goal of CSI inference, two reasonable assumptions are made, which describe the capability boundary of CSI inference, i.e., parameters that can be inferred and those cannot based on realistic rationality. Specifically,
	\begin{itemize}
		\item 
		The phase of an arrival MPC, i.e., $\phi_{\gamma}$, is random (i.i.d. among MPCs) and cannot be inferred. This is a practical and realistic consideration given the fact that the phase of an electromagnetic wave shifts dramatically even with a slight movement of the terminal (several wavelengths).
		\item
		The observable scatterers seen at the remote site, i.e., $S_{\mathsf{rm}}(\gamma)$, $r_{\mathsf{max},\mathsf{rm}}$, cannot be inferred based on the observation at local site which provides little information about whether an MPC is obstructed seen at the remote site.  Instead, this information can be obtained by using a relatively infrequent probing signals by the remote site given the fact that the scattering environment is constant inside the stationarity region (typical size of tens of meters in urban areas).\footnote{The requirement for pilot signals to probe the MPC distributions is mainly assumed for theoretical analysis; based on the deep learning implementation, these pilots are usually unnecessary to obtain a good CSI inference performance.}
	\end{itemize}
	
	Based on these assumptions, we focus on analyzing the inference performance with respect to the AoA at the remote site $\theta_{\mathsf{rm}}$ to obtain theoretical results. The end goal is to derive the CRLB of the estimation of $\theta_{\mathsf{rm}}$, towards which we first solve the inverse problem of $h_{\mathsf{lc},i} = \operatorname{f}\left(\theta_{\mathsf{lc}}, S_{\mathsf{lc}}(\gamma), \phi_{\mathsf{lc},\gamma}, r_{\mathsf{max},\mathsf{lc}}\right)$, and then relates to the AoA of $\theta_{\mathsf{rm}}$ based on geometry. Based on the first assumption, the phase is random and hence the received signal in \eqref{int_res} can be viewed as a zero-mean circular complex Gaussian process (assuming a large number of scatterers), whose probability distribution function (pdf) is completely characterized by its covariance matrix (sufficient statistics)
	\begin{equation}
	\label{cyc}
	\C_y = \mathbb{E}[\y \y^{\mathsf{H}}] = P_{\textrm{tx}} \mathbb{E}\left[\h \h^{\mathsf{H}}\right] + \C_n,
	\end{equation}
	where $\C_n$ is the noise covariance, and its estimation as
	\begin{equation}
	\hat{\C}_y = \frac{1}{K}\sum_{k=1}^{K} \y(k)\y^{\mathsf{H}}(k),
	\end{equation}
	where $K$ is the number of sampling points, $\y(k)$ denotes the $k$-th sample, and furthermore 
	\begin{iarray}
		\label{rho}
		&& \left\{\mathbb{E}\left[\h \h^{\mathsf{H}}\right]\right\}_{ml} = \mathbb{E}\left[h_m h_l^{\sf{H}}\right] \nonumber\\
		&=&  \int_{-\gamma_{\mathsf{max}}}^{\gamma_{\mathsf{max}}}  S(\gamma) \exp{\left(-\frac{j2\pi \delta m}{\lambda}  \cos{(\gamma+\theta)} + j \phi_{\gamma} \right)} \dx \gamma \nonumber\\
		&& \times \int_{-\gamma_{\mathsf{max}}}^{\gamma_{\mathsf{max}}}  S(\gamma) \exp{\left(\frac{j2\pi \delta l}{\lambda}  \cos{(\gamma+\theta)} - j \phi_{\gamma} \right)} \dx \gamma \nonumber\\
		&\overset{(a)}{=}&  \int_{-\gamma_{\mathsf{max}}}^{\gamma_{\mathsf{max}}}  S^2(\gamma) \exp{\left(-\frac{j2\pi \delta (m-l)}{\lambda}  \cos{(\gamma+\theta)}  \right)} \dx \gamma \nonumber\\
		&\overset{(b)}{=}&  \exp{\left(-\frac{j2\pi \delta (m-l)}{\lambda}  \cos{\theta} \right)}\nonumber\\
		&& \times \int_{-\gamma_{\mathsf{max}}}^{\gamma_{\mathsf{max}}}  S^2(\gamma) \exp{\left(\frac{j2\pi \delta (m-l)}{\lambda}  \sin{\theta} \gamma \right)} \dx \gamma \nonumber\\
		&\overset{(c)}{=}&  \exp{\left(-\frac{j2\pi \delta (m-l)}{\lambda}  \cos{\theta} \right)} \nonumber\\ 
		&& \times \mathcal{F}^{-1}\left\{S^2(\gamma)\mathsf{rect}\left(\frac{\gamma}{2 \gamma_{\mathsf{max}}}\right)\Bigg|_{f=\gamma}\right\}\Bigg|_{t=\frac{(m-l)\delta \sin{\theta}}{\lambda}}.
	\end{iarray}
	The equality of $(a)$ is based on the fact that the arrival phases of MPCs are assumed i.i.d. and hence the cross terms in the integral are averaged out. The equality of $(b)$ is based on the approximation that $r_{\mathsf{max}}$ is small and hence
	\begin{equation}
	\sin{\gamma} \approx \gamma \textrm{ and } \cos{\gamma} \approx 1.
	\end{equation}
	The equality of $(c)$ is obtained by employing the inverse Fourier transform and $\mathsf{rect}(\cdot)$ denotes the rectangular function. The channel covariance matrix can be obtained with each entry given in \eqref{rho}. Thereby, we are ready to derive the CRLB of channel inference. Given the observations at the local site $\C_{y_\mathsf{lc}}$ in \eqref{cyc}, the log-likelihood function can be written as
	\begin{equation}
	\mathcal{L}(\z) = -K \log\left|\C_{y_\mathsf{lc}}\right| - K \mathsf{tr} \left[\C^{-1}_{y_\mathsf{lc}} \hat{\C}_{y_\mathsf{lc}}\right] + \mathsf{const}.
	\end{equation}
	where $\z=\left\{S(\gamma),\gamma_\mathsf{max},\theta_{\mathsf{lc}}\right\}$. Its derivative can be calculated as
	\begin{iarray}
		\mathsf{d} \mathcal{L}(\z) = -K \mathsf{tr} \left[\C^{-1}_{y_\mathsf{lc}} -  \C^{-1}_{y_\mathsf{lc}} \hat{\C}_{y_\mathsf{lc}} \C^{-1}_{y_\mathsf{lc}} \right] \mathsf{d} \C_{y_\mathsf{lc}}.
	\end{iarray}
	The Fisher information matrix {FIM} is given by
	\begin{iarray}
		\label{fim}
		\left\{\J_{\z}\right\}_{ij} &=& -\mathbb{E}\left[\frac{\partial^2 \mathcal{L}(\z)}{\partial \z_i \partial \z_j}\right] \nonumber\\
		&=& K \mathbb{E} \left[ \frac{\partial \left(\mathsf{tr} \left[\I_{M_\mathsf{lc}} -  \C^{-1}_{y_\mathsf{lc}} \hat{\C}_{y_\mathsf{lc}}  \right] \C^{-1}_{y_\mathsf{lc}} \D_i\right)}{\partial \z_j}\right] \nonumber\\
		&=& K \mathsf{tr} \left[ \C^{-1}_{y_\mathsf{lc}}  \mathbb{E}\left[\hat{\C}_{y_\mathsf{lc}}\right]  \C^{-1}_{y_\mathsf{lc}}  \D_j \C^{-1}_{y_\mathsf{lc}} \D_i \right] \nonumber\\
		&& + K   \mathsf{tr} \left[\I_{M_\mathsf{lc}} -  \C^{-1}_{y_\mathsf{lc}} \mathbb{E} \left[ \hat{\C}_{y_\mathsf{lc}}\right]\right]  \frac{\partial \left( \C^{-1}_{y_\mathsf{lc}} \D_i\right)}{\partial \z_j} \nonumber\\
		&\overset{(a)}{=}& K \mathsf{tr} \left[ \C^{-1}_{y_\mathsf{lc}} \D_j \C^{-1}_{y_\mathsf{lc}} \D_i \right],
	\end{iarray}
	where we use $\mathbb{E} \left[ \hat{\C}_{y_\mathsf{lc}}\right] = {\C}_{y_\mathsf{lc}}$ in equality $(a)$, and $\D_i=\frac{\partial \C_{y_\mathsf{lc}}}{\partial z_i}$. Based on \eqref{fim}, the CRLB of AoA $(\theta_\mathsf{lc})$ and terminal distance $D_\mathsf{lc}$ can be obtained by extracting the diagonal entries of the inverse FIM.
	
	The CRLB of $\theta_\mathsf{rm}$ can be therefore obtained in the following. Without loss of generality, assume the coordinates of the local site and remote site are $(0,0)$ and $(D_0\cos{\theta_0},D_0\sin{\theta_0})$ in a two-dimensional Cartesian coordinate system, respectively. Based on $\theta_\mathsf{lc}$ and $D_\mathsf{lc}$, the coordinate of the terminal can be expressed in two ways as follows:
	\begin{equation}
	x_\mathsf{t} = D_\mathsf{lc}\cos{\theta_\mathsf{lc}} = D_0\cos{\theta_0} + D_\mathsf{rm}\cos{\theta_\mathsf{rm}},
	\end{equation}
	\begin{equation}
	y_\mathsf{t} = D_\mathsf{lc}\sin{\theta_\mathsf{lc}} = D_0\sin{\theta_0} + D_\mathsf{rm}\sin{\theta_\mathsf{rm}}.
	\end{equation}
	The AoA at the remote site can be hence represented by $D_\mathsf{lc}$ and $D_\mathsf{rm}$ as follows
	\begin{equation}
	\label{theta_rm}
	\theta_\mathsf{rm} = \arctan\left(\frac{D_\mathsf{lc}\sin{\theta_\mathsf{lc}}-D_0\sin{\theta_0}}{D_\mathsf{lc}\cos{\theta_\mathsf{lc}}-D_0\cos{\theta_0}}\right).
	\end{equation}
	The CRLB of $\theta_\mathsf{rm}$ can be expressed as:
	\begin{iarray}
		\label{crb}
		\mathsf{CRB}(\theta_\mathsf{rm}) &=& \left(\frac{\partial \theta_\mathsf{rm}}{\partial D_{\mathsf{lc}}}\right)^2 \mathsf{CRB}(D_{\mathsf{lc}}) + \left(\frac{\partial \theta_\mathsf{rm}}{\partial \theta_{\mathsf{lc}}}\right)^2 \mathsf{CRB}(\theta_{\mathsf{lc}}) \nonumber\\
		&=& \frac{D^2_0\sin^2{(\theta_0-\theta_{\mathsf{lc}})}}{D^4_\mathsf{rm}}\mathsf{CRB}(D_{\mathsf{lc}}) \nonumber\\
		&& +\frac{D_{\mathsf{lc}}^2 
		\left( D_{\mathsf{lc}} - D_0\cos{(\theta_0-\theta_{\mathsf{lc}})}\right)^2}{D_\mathsf{rm}^4}\mathsf{CRB}(\theta_{\mathsf{lc}}),
	\end{iarray}
	where $\mathsf{CRB}(D_{\mathsf{lc}})$ and $\mathsf{CRB}(\theta_{\mathsf{lc}})$ are obtained from \eqref{fim}.
	
	The calculation of $\D_i$ in \eqref{fim} depends on the channel model, e.g., scattering distribution inside the ring. It seems elusive to calculate closed-form expressions for general models. To gain some insights, we consider several special cases in the following where the angular spreads $\gamma_{\mathsf{max},\mathsf{lc}}$ and $\gamma_{\mathsf{max},\mathsf{rm}}$ are small and approach zero, i.e., a LoS MPC only.
	\subsection{Special Case: LoS Scenario and One Local Site}
	Considering the LOS case, $\gamma_\mathsf{max}=0$, $S(\gamma)=\Delta(\gamma)$ $(\Delta(x)=0,\,\forall x \neq 0,\,\textrm{ and }\int_{-\infty}^{\infty}\Delta(x)=1)$, and hence 
	\begin{equation}
	\y_{y_\mathsf{lc,LOS}} = \rho_\mathsf{lc} \exp{(j\phi)}  \e  + \n,
	\end{equation}
	where $\rho_\mathsf{lc}=\sqrt{P_\textrm{tx}M_\mathsf{lc}} \frac{\lambda}{4\pi D_\mathsf{lc}}$, $\left\{\e\right\}_i=\frac{e^{\frac{-j2\pi i \delta }{\lambda}  \cos{\theta_\mathsf{lc}} }}{\sqrt{M_\mathsf{lc}}}$. Due to the fact that there is only one LOS MPC, the Gaussianity of the array response is lost; specifically the first term on the right-hand side of the equation is deterministic and therefore we have 
	\begin{equation}
	\y_{y_\mathsf{lc,LOS}} \sim \mathcal{CN}(\rho_\mathsf{lc} \exp{(j\phi)}  \e,\sigma^2 \I_{M_\mathsf{lc}}).
	\end{equation}
	A small modification to \eqref{fim} is required to account for non-zero mean, which reads $(\z=\left[\rho_\mathsf{lc},\tau_\mathsf{lc},\phi\right])$
	\begin{iarray}
		\left\{\J_{\z}\right\}_{ij} &=& K \mathsf{tr} \left[ \C^{-1}_{y_\mathsf{lc}} \D_j \C^{-1}_{y_\mathsf{lc}} \D_i \right] \nonumber\\
		&& + \frac{2K}{\sigma^2} \left(\frac{\partial \u}{\partial z_j} \frac{\partial \u}{\partial z_i} + \frac{\partial \v}{\partial z_i} \frac{\partial \v}{\partial z_j}\right),
	\end{iarray}
	where in this case $\m=\rho_\mathsf{lc} \exp{(j\phi)}  \e \triangleq \u + j \v$, $\C_{y_\mathsf{lc}}=\sigma^2\I_{M_\mathsf{lc}}$ and therefore $\D_i=\boldsymbol{0}$, $\forall i$.	Denote $\tau_\mathsf{lc} \triangleq -\frac{2\pi \delta \cos{\theta_{\mathsf{lc}}}}{\lambda} $, then
	\begin{iarray}
		\left\{\J_{\z}\right\}_{11} &=& \frac{2K}{\sigma^2} \e^\mathsf{H} \e = \frac{2K}{\sigma^2}, \nonumber\\
		\left\{\J_{\z}\right\}_{12} &=& \left\{\J_{\z}\right\}_{21} = 	\left\{\J_{\z}\right\}_{13} = 	\left\{\J_{\z}\right\}_{31} = 0, \nonumber\\
		\left\{\J_{\z}\right\}_{22} &=& \frac{2K\rho_\mathsf{lc}^2}{\sigma^2} \left(\frac{\partial \e}{\partial \tau_\mathsf{lc}}\right)^{\mathsf{H}} \frac{\partial \e}{\partial \tau_\mathsf{lc}} = \frac{K\rho_\mathsf{lc}^2(M_\mathsf{lc}-1)(2M_\mathsf{lc}-1)}{3\sigma^2}, \nonumber\\
		\left\{\J_{\z}\right\}_{23} &=& \left\{\J_{\z}\right\}_{32} = \frac{M_\mathsf{lc}-1}{\sigma^2} K \rho_\mathsf{lc}^2, \nonumber\\
		\left\{\J_{\z}\right\}_{33} &=& \frac{2 K \rho_\mathsf{lc}^2}{\sigma^2}.
	\end{iarray}
	The CRLBs can be readily derived as
	\begin{iarray}
	\mathsf{CRB}(\rho_\mathsf{lc}) &=& \left\{\J_{\z}^{-1}\right\}_{11} = \frac{\sigma^2}{2K}, \nonumber\\
	\mathsf{CRB}(\tau_\mathsf{lc}) &=& \left\{\J_{\z}^{-1}\right\}_{22} = \frac{6 \sigma^2}{K\rho_\mathsf{lc}^2(M^2_\mathsf{lc}-1)M_\mathsf{lc}},
	\end{iarray}
	respectively. Similar with \eqref{crb}, we can then obtain the CRLBs of $D_\mathsf{lc}$ and $\theta_\mathsf{lc}$ as:
	\begin{iarray}
	\mathsf{CRB}(D_\mathsf{lc}) &=& \frac{8\pi^2 D_\mathsf{lc}^4\sigma^2}{\lambda^2 K P_\textrm{tx} M_\mathsf{lc}}, \nonumber\\
	\mathsf{CRB}(\theta_\mathsf{lc}) &=& \frac{24\sigma^2D_\mathsf{lc}^2}{K M_\mathsf{lc}(M_\mathsf{lc}^2-1) P_\textrm{tx} \delta^2\textup{sin}^2{\theta_\mathsf{lc}}}.
	\end{iarray}
	Denote the effective receive signal-to-noise ratio as $\mathsf{SNR} = \frac{P_\textrm{tx}\left(\frac{\lambda}{4 \pi D_\mathsf{lc}}\right)^2}{\sigma^2}$, and $\delta=\lambda/2$, then
	\begin{iarray}
		\label{crb_los_2sites}
		\mathsf{CRB}_1(\theta_\mathsf{rm}) &=& \frac{D_{\mathsf{lc}}^2  }{D_\mathsf{rm}^4} \frac{1}{K\mathsf{SNR}} \underbrace{\left(\frac{c_1 D_0^2 \sin^2{(\theta_0-\theta_{\mathsf{lc}})}}{M_\mathsf{lc}}\right.}_{\mathcal{M}_1} \nonumber\\
		&& + \underbrace{\left.\frac{c_2 \left( D_{\mathsf{lc}} - D_0\cos{(\theta_0-\theta_{\mathsf{lc}})}\right)^2 }{{M_\mathsf{lc}(M_\mathsf{lc}^2-1)}\sin^2{\theta_\mathsf{lc}}}\right)}_{\mathcal{M}_2},
	\end{iarray}
	where $c_1=\frac{1}{2}$, $c_2=\frac{6}{\pi^2}$. It is noted that
	\begin{iarray}
	\mathcal{M}_1 \sim \frac{1}{M_\mathsf{lc}},\textrm{ and }\mathcal{M}_2 \sim \frac{1}{M^3_\mathsf{lc}},
	\end{iarray}
	and hence
	\begin{iarray}
	\mathsf{CRB}_1(\theta_\mathsf{rm}) \sim \frac{1}{M_\mathsf{lc}}.
	\end{iarray}
	Therefore the inference performance bottleneck is at $\mathcal{M}_1$, i.e., the inference error related to distance estimation which is reciprocal with $M_\mathsf{lc}$. On the other hand, the inference error associated with AoA estimation scales inversely with $M_\mathsf{lc}^3$. Based on this observation, the performance can be improved by leveraging CSI at multiple sites and correspondingly multiple AoAs to make the inference. We present the following CRLB analysis which accounts for two separate local sites (with known CSI) to infer the CSI at a remote cite.
	\subsection{Special Case: LoS Scenario with Known CSI at Two Sites}
	In this subsection, we will show that by using the AoAs at two geographically separated sites, the inference error can be significantly reduced; such a scenario presents itself in densely deployed cellular systems and, moreover, the cost of acquiring CSI at one other site is affordable. 
	
	Inheriting the denotations in the last subsection, denote the location of the other local site as, without loss of generality, $(D_\mathsf{lc}^\prime,0)$ with $D_\mathsf{lc}^\prime > 0$, and denote the AoA at the other local site as $\theta_\mathsf{lc}^\prime$. The AoA at the remote site, i.e., $\theta_\mathsf{rm}$, can be expressed by 
	\begin{equation}
	\theta_\mathsf{rm} = \arctan\left(\frac{D_\mathsf{lc}^\prime \sin\theta_\mathsf{lc}^\prime\sin{\theta_\mathsf{lc}}-D_0\sin{\theta_0} \sin\left(\theta_\mathsf{lc}^\prime-\theta_\mathsf{lc}\right)}{D_\mathsf{lc}^\prime \sin\theta_\mathsf{lc}^\prime \cos{\theta_\mathsf{lc}}-D_0\cos{\theta_0} \sin\left(\theta_\mathsf{lc}^\prime-\theta_\mathsf{lc}\right)}\right).
	\end{equation}
	The CRLB of $\theta_{\mathsf{rm}}$ with known CSI at two local sites is
	\begin{iarray}
		\label{crb_los}
		\mathsf{CRB}_2(\theta_\mathsf{rm}) &=& \frac{6 D_{\mathsf{lc}}^2  }{\pi^2D_\mathsf{rm}^4 {M_\mathsf{lc}(M_\mathsf{lc}^2-1)} } \frac{1}{K\mathsf{SNR}} \nonumber\\
		&& \times \frac{\omega_1 + \omega_2}{\sin^2{\left(\theta_\mathsf{lc}-\theta^\prime_\mathsf{lc}\right)}\sin^2{\theta_\mathsf{lc}}},
	\end{iarray}
	where 
	\begin{iarray}
	\omega_1 &=& D_0^2 \sin^2{\theta_\mathsf{lc}} \sin^2(\theta_0-\theta_\mathsf{lc}) \nonumber\\
	\omega_2 &=& \sin^2{\theta_\mathsf{lc}^\prime} \left(D_\mathsf{lc}^\prime \sin{\theta_\mathsf{lc}^\prime} - D_0 \sin(\theta_\mathsf{lc}^\prime-\theta_0) \right)^2.
	\end{iarray}
	It follows that
	\begin{iarray}
	\mathsf{CRB}_2(\theta_\mathsf{rm}) \sim \frac{1}{M^3_\mathsf{lc}}.
	\end{iarray}
	\begin{remark}
	While this paper focuses on inference based solely on spatial domain signals, one direction that is certainly worth studying is the CSI inference performance incorporating wideband signals such that the time-of-arrival (ToA) can also be estimated. $\hfill\square$
	\end{remark}
	\section{CSI Inference Based on DNN}
	\label{sec_dnn}
	\begin{figure}[!t]
    	\centering
    	\includegraphics[width=0.46\textwidth]{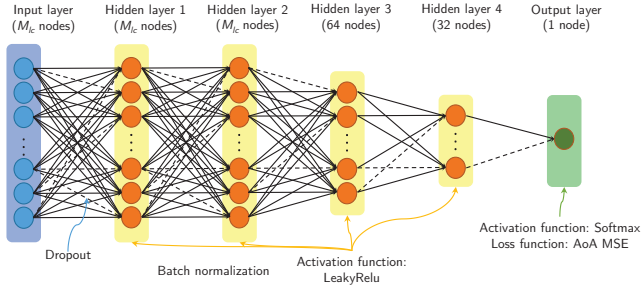}
    	\caption{The proposed DNN architecture.}
    	\label{Fig_dnn}
    \end{figure}
    CSI inference in real-world signal propagation scenarios is extremely difficult and intractable, and moreover existing model-based approaches are inadequate to address this issue. Therefore, inspired by the recent advance in deep learning field, we develop a DNN to accomplish this task. A DNN with $4$ hidden layers is adopted. The sizes of layers are $M_\mathsf{lc}$, $M_\mathsf{lc}$, $64$ and $32$, respectively. The non-linear function for the hidden layers is $\mathsf{LeakyReLU}$, while the non-linear function for the output layer is $\mathsf{sigmoid}$. The input features are quantized CSI in the angular domain (take modulus and logarithm) which are normalized by Z-score normalization method while the output is normalized AOA (AOA divided by $\pi$). Here the prescribed codebook is the discrete-Fourier-transform (DFT) codebook, and the size of the codebook is $M_\mathsf{lc}$. 

    In order to avoid overfitting, we apply dropout \cite{dropout} in our neural network and the keep probability is set to be $0.7$. The cost function of the neural network is mean-squared-error (MSE) of AOA and the MSE is also our performance metric. The Adam gradient-based optimizer is used with step size of $10^{-4}$. We randomly divide data set into two parts for training $(90\%)$ and test $(10\%)$. The trained model runs for $10$ times and the average is taken.

    \section{Simulation Results}
    \label{sec_nr}
\begin{figure}[!t]
    	\centering
    	\includegraphics[width=0.46\textwidth]{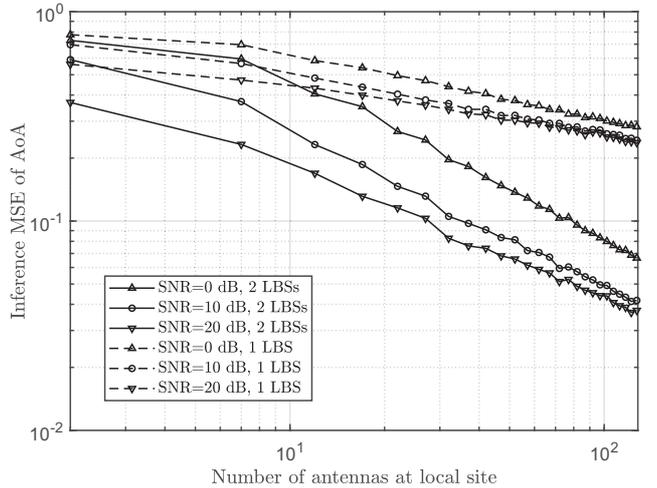}
    	\caption{DNN-based CSI inference performance.}
    	\label{Fig_dnninfer}
    \end{figure}

\begin{figure}[!t]
    	\centering
    	\includegraphics[width=0.46\textwidth]{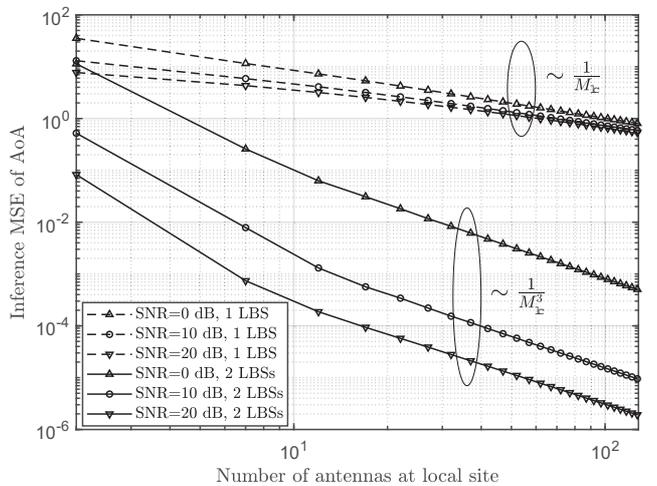}
    	\caption{CRLB of inference error for one-ring channel model.}
    	\label{Fig_crlbinfer}
    \end{figure}

    In Fig. \ref{Fig_dnninfer} and \ref{Fig_crlbinfer}, we present simulation results based on the one-ring channel model. The two local BSs (LBSs) are located at $(-100,0)$ and $(100,0)$ (the distance unit is meter), the remote BS (RBS) whose CSI is to be inferred is at $(0,50)$ and the terminal is randomly located on a semi-disk (to avoid AoA phase ambiguity of ULA) centered at the RBS with a radius of $50$~m (not allowed to be within $5$~m to the RBS). The DNN is as described in Section \ref{sec_dnn}. The results, which are plotted on logarithmic scale, indicate that the scaling law derived for the LoS scenario, i.e., the CRLB scales with $1/M_\mathsf{lc}$ and $1/M_\mathsf{lc}^3$ with one LBS and two LBSs respectively, is upheld for the one-ring channel model. More importantly, it is found that the inference performance by DNN also exhibits such behaviour, while the exact scaling factors may vary due to the unclear noise structure of the black-box of DNN. We should emphasize that it may be unfair to compare the performance of DNN and CRLB since DNN is applied to a much more general scenario while the CRLB is concerned with the optimal channel estimator in this specific channel model. However, the insights given by analyzing the CRLB are still useful as it is shown that the qualitative results are consistent between the DNN performance and analytical CRLB.
    \begin{figure}[!t]
        \centering
        \subfigure{
        \includegraphics[width=0.19\textwidth]{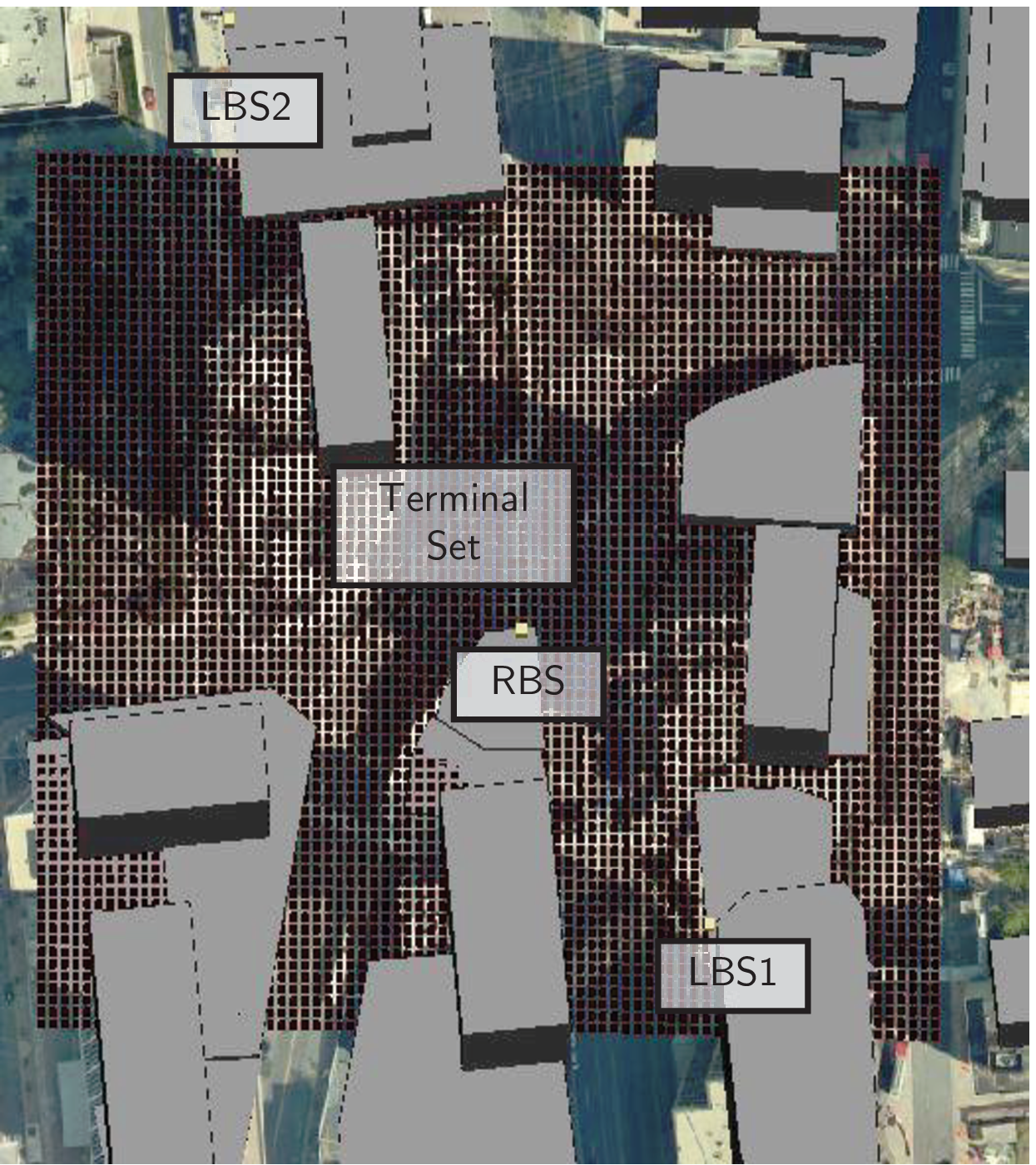}
        \label{Fig_raymodel}}
        \subfigure{
        \includegraphics[width=0.26\textwidth]{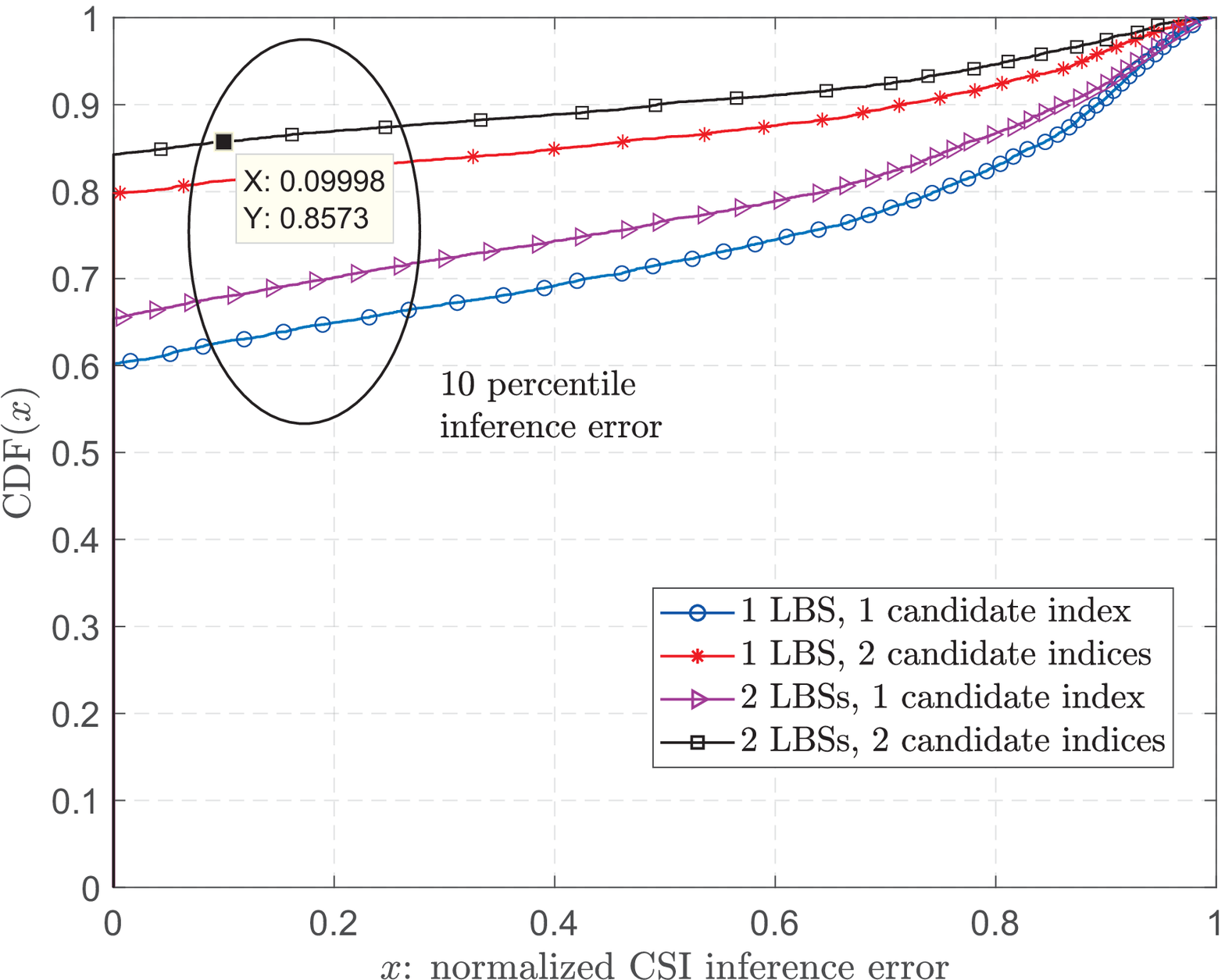}
        \label{Fig_rayinfer}}
        \caption{DNN-based CSI inference performance with a realistic ray-tracing channel simulator.}
        \label{Fig_ray}
    \end{figure} 
    
    The DNN-based approach is tested in a more realistic scenario in Fig. \ref{Fig_ray}. We use the Wireless Incite\textsuperscript{\textregistered} ray-tracing channel simulator to generate the CSI. The system layout is depicted in Fig. \ref{Fig_raymodel}. The antenna arrays are two-dimensional with sizes of $4 \times 100$ and $4 \times 20$ for LBS and RBS respectively; both arrays are located with heights of $20$~m. The carrier frequency is $28$~GHz and the system bandwidth is $10$~MHz. $20000$ samples are obtained for training and test. In this case where the propagation channel may be non-LoS, the beamforming vector (DFT codebook) with the largest power, instead of the AoA which is not well-defined in this case, is the inference objective; therefore the output layer is modified to have the same number of neurons as the number of codewords in the DFT codebook, and the loss function is changed to the cross-entropy function compared with the AoA MSE in Fig. \ref{Fig_dnn}. The performance metric is still the inference error normalized in the DFT codebook (to values within $[0,1]$). Concretely, the normalized inference error is defined as
    \begin{equation}
        e \triangleq 1-\frac{\hat{\h}^\dag\h}{\h_{\mathsf{opt}}^\dag\h},
    \end{equation}
    where $\hat{\h}$ is the inferred CSI from the DFT codebook, $\h_{\mathsf{opt}}$ is the optimal DFT beamforming vector in the codebook and the CSI obtained by the ray-tracing simulator is denoted by $\h$. It is observed from Fig. \ref{Fig_rayinfer} that about $85\%$ of the time, the optimal beamforming vector of RBS can be inferred with an error less than $10\%$ based on observations at LBSs in a realistic scenario.  
    \section{Conclusions}
    \label{sec_con}
    The remote CSI inference problem is considered both theoretically and practically. To establish the relevance between CSI of LBS and RBS, we use the physical propagation environment as the bridge, however the inference problem is identified as ill-posed with insufficient observation capability. With proper simplifications by adopting the one-ring channel model, we transform the inference problem to a parameter extraction problem and derive the CRLB thereof. For a special LoS case, the closed-form expressions of CRLB can be obtained, showing that the CRLB scales inversely with $M_\mathsf{lc}$ and $M_\mathsf{lc}^3$ by using one and two LBSs respectively. An improved DNN with dropout to thwart overfitting and a deeper network is proposed, showing robustly good performance in real-world scenarios. It is found by simulations that the theoretical CRLB analysis, although significantly simplified, provides insightful performance observations and meaningful implications.
	\section*{Acknowledge}
	This work is sponsored in part by the Nature Science Foundation of China (No. 61701275, No. 91638204, No. 61571265, No. 61621091), the China Postdoctoral Science Foundation, and Intel Collaborative Research Institute for Mobile Networking and Computing.
	\bibliography{sm}
	\bibliographystyle{IEEEtran}
\end{document}